\documentclass[a4paper,11pt]{article}
\usepackage{pos}
\usepackage{gensymb}
\usepackage{subfig}

\title{A machine learning approach to axion-like particle searches in CTAO observations of blazars}
\ShortTitle{Machine learning approach to ALP searches in blazars}

\author[a,b]{Francesco Schiavone}
\author*[b]{Leonardo Di Venere}
\author[a,b]{Francesco Giordano}
\onbehalf{for the CTAO Consortium}

\affiliation[a]{Dipartimento Interateneo di Fisica ``Michelangelo Merlin'', Università degli Studi di Bari, Via Amendola 173, 70125, Bari, Italy}
\affiliation[b]{INFN Sezione di Bari, Via Amendola 173, 70125, Bari, Italy}

\emailAdd{francesco.schiavone@ba.infn.it}
\emailAdd{leonardo.divenere@ba.infn.it}

\abstract{Axion-like particles (ALPs) are a common prediction of several extensions of the Standard Model of particle physics and could be detected through their coupling to photons, which enables ALP-photon conversions in external magnetic fields. This conversion could lead to two distinct signatures in gamma-ray spectra of blazars: a superimposition of energy-dependent "wiggles" on the spectral shape, and a hardening at high (multi-TeV) energies, due to the ALP beam eluding absorption by the extragalactic background light (EBL). The enhanced energy resolution of the Cherenkov Telescope Array Observatory (CTAO) with respect to present ground-based gamma-ray telescopes makes it an ideal instrument to probe such phenomena. In this contribution, we explore a different approach based on the use of machine learning (ML) classifiers and compare it to the standard method. Our preliminary results suggest that both techniques yield consistent results, with the ML-based method offering comparable or even slightly broader coverage, potentially extending the CTAO sensitivity beyond existing constraints.}

\ConferenceLogo{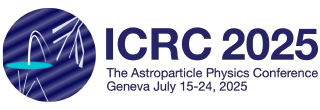}

\FullConference{39th International Cosmic Ray Conference (ICRC2025)\\
 15–24 July 2025\\
Geneva, Switzerland\\}

\newcommand{\gag}{g_{a\gamma}}
\newcommand{\Pgg}{P_{\gamma\gamma}}

\begin{document}
\maketitle

\section{Introduction}
Axion-like particles (ALPs) are hypothetical, pseudo-scalar particles whose existence is predicted by several extensions of the Standard Model. A fundamental property of these particles, whose prototype is represented by the QCD axion introduced by Peccei, Quinn, Weinberg and Wilczek \cite{pecceiquinn1977, weinberg1978, wilczek1978}, is an effective coupling to photons, parametrized by the coupling constant $\gag$. The existence of this coupling is essential for the experimental detection of ALPs, as it allows for ALP-photons oscillations to occur in the presence of external magnetic fields.

A particular manifestation of this effect might be observed in the relativistic jets of blazars, a class of active galactic nuclei (AGNs) featuring powerful relativistic jets forming small angles ($\theta\lesssim20\degree$) with the line of sight from Earth. The gamma-ray photons emitted by the blazar could convert to ALPs inside the jets, which typically feature strong magnetic fields over pc scales, traverse the extragalactic background light (EBL) and finally convert back to photons in the Milky Way magnetic field. This process would lead to distinct features in the observed gamma-ray spectra of blazars, so-called ``wiggles''. Additionally, a flux enhancement above TeV energies would be observed, as the very-high-energy (VHE) gamma rays converting to ALPs would escape EBL absorption \cite{deangelis2013}. These effects are summarised by defining an energy-dependent photon survival probability $\Pgg(E)$, which relates the observed gamma-ray flux $\phi_{\rm obs}$ to the intrinsic one $\phi_{\rm int}$ as 
\begin{equation}
\label{eq:obs-spectrum-alp}
    \phi_{\rm obs}(E)=\phi_{\rm int}(E)\Pgg(E)\,.
\end{equation}

The observation of astrophysical gamma-ray sources has allowed several upper limits to be placed over the two-dimensional parameter space spanned by the ALP mass $m_a$ and the $\gag$ coupling constant \cite{batkovic2021}. A typical approach for computing these constraints is to estimate the significance for excluding the ALP hypothesis by defining a test statistic (TS) based on the likelihood ratio \cite{abdalla2021, abe2024}. In this contribution, we present instead a new method based on the properties of machine learning (ML) classifiers. We apply this method to simulated observations of selected sources, to be performed with the future Cherenkov Telescope Array Observatory (CTAO\footnote{\href{https://www.ctao.org/}{https://www.ctao.org/}}).

\section{Source modelling and simulation}
In this work we have selected two well-known high synchrotron-peaked BL Lac (HBL) objects, namely Markarian 501 (Mrk 501) and PKS 2155$-$304. The choice of these sources was motivated by their significant TeV emission, as well as by their inclusion in the CTAO AGN Key Science Project \cite{ctascience2018}, which will guarantee the availability of high-quality data during the first years of operation of the Observatory.

\begin{figure}
    \centering
    \includegraphics[width=0.6\linewidth]{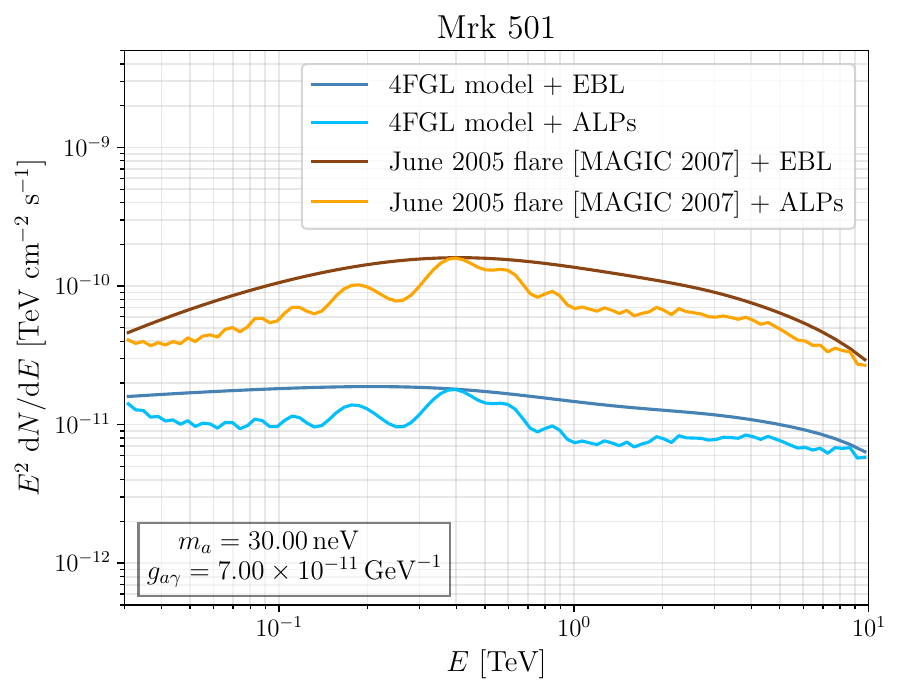}
    \caption{ALP effects on the baseline and flaring spectral shapes chosen for Mrk 501, one of the sources considered in this work.}
    \label{fig:alp-spectra}
\end{figure}

For each source, we considered a baseline state from the 4th \textit{Fermi}-LAT Source Catalog Data Release 4 (4FGL-DR4) \cite{4fgl-dr3, 4fgl-dr4}, as well as a reference flaring state from previous TeV observations \cite{albert2007, acciari2011, aharonian2009}. Both sets of models were assumed to be valid in the $30\,{\rm GeV}$--$10\,{\rm TeV}$ energy range. We computed the photon survival probability $\Pgg$ for each source by using the gammaALPs v0.3.0\footnote{\href{https://github.com/me-manu/gammaALPs}{https://github.com/me-manu/gammaALPs}} \cite{meyer2021} Python package. The jet magnetic field parameters from ref. \cite{potter2015} were included, as well as the free electron density normalization from ref. \cite{tavecchio2010}. Additionally, EBL absorption and the Milky Way's magnetic field were modeled according to refs. \cite{dominguez2011} and \cite{jansson2012}, respectively. The spectral models used for Mrk 501, with and without ALP effects, are shown as an example in Figure \ref{fig:alp-spectra}.

We simulated CTAO observations using the Gammapy v1.0.1 Python package \cite{gammapy2023, gammapy-1.0.1}. We used the publicly available prod5 instrument response functions (IRFs) \cite{ctao_irfs_prod5} for the initial configurations of the Northern Array for Mrk 501 and Mrk 421 and of the Southern Array for PKS 2155$-$304. We considered 50 hours of observation for the baseline states and 5 hours for the flaring states, assuming a zenith distance of 20 degrees while averaging over the azimuth angle. The observations were simulated in the ON-OFF mode with a $0.7\degree$ offset and an acceptance ratio of $1/3$, without including ALP effects. Following ref. \cite{abdalla2021}, we used 40 energy bins per decade, and we employed Asimov datasets \cite{cowan2011} in order to avoid computing a mean sensitivity from many different simulations.

\section{Method}
We consider a grid of $10\times10$ logarithmically spaced points over the ALP parameter space, with $m_a\in[0.1, 1000]\,{\rm neV}$ and $\gag\in[0.03,7]\times 10^{-11}\,{\rm GeV}^{-1}$. For each $(m_a, \gag)$ point and for each source, we define a ML classifier based on the XGBoost algorithm \cite{chen2016}. The classifier is trained to distinguish between 2000 simulated datasets with and without ALP effects, using the excess photon counts above background in each bin (normalized
between 0 and 1) as features. The performance of the trained ``classifier grid'' can be evaluated by computing the resulting classifier accuracy over the ALP parameter space. 

\begin{figure}
    \centering
    \subfloat[\label{fig:example-ts-distribution-ml-good}]{
        \includegraphics[width=.485\textwidth]{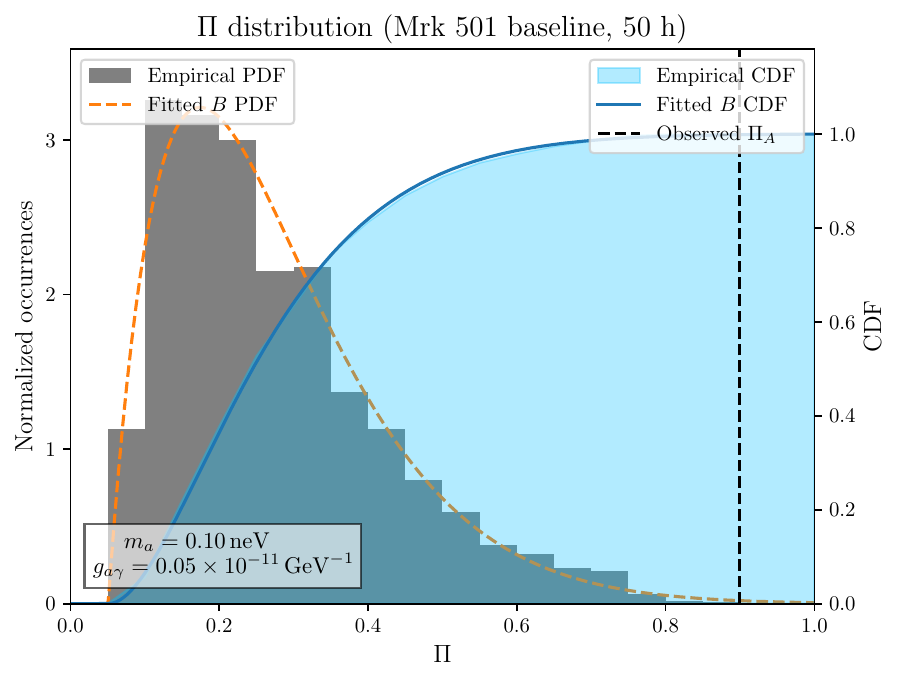}
        }
    \hfill
    \subfloat[\label{fig:example-ts-distribution-ml-bad}]{
        \includegraphics[width=.485\textwidth]{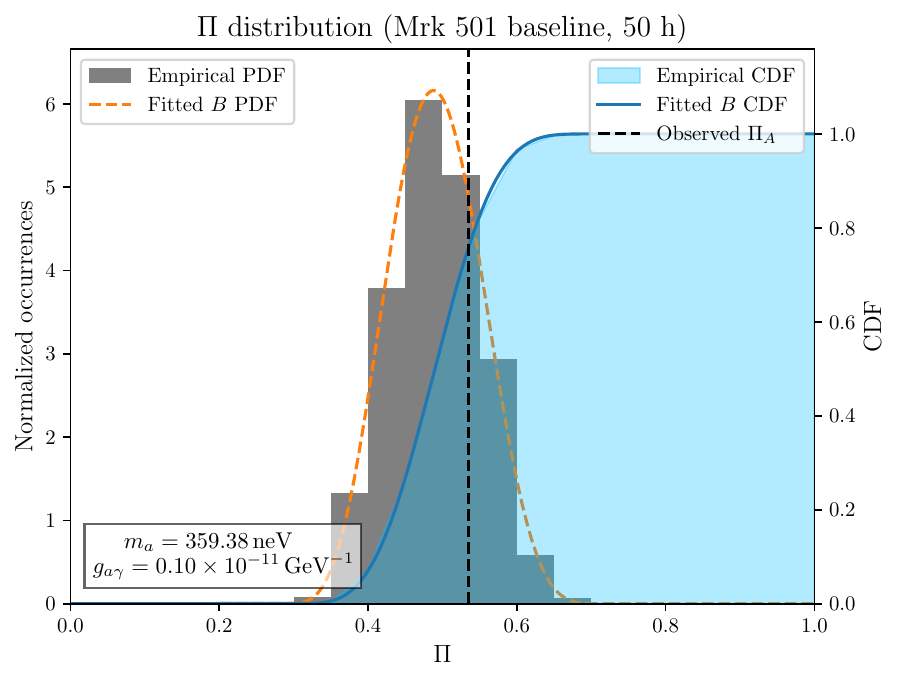}
        }
       
    \caption{Example distributions of the $\Pi$ statistic defined in eq. \eqref{eq:ts-ml}, comparing good and suboptimal performance of a classifier algorithm in different regions of the ALP parameter space.}
    \label{fig:example-ts-distribution-ml}
\end{figure}

Each classifier returns a probability $p_{\rm ALP}(m_a,\gag|D)$ that a certain dataset $D$ is ALP-like , i.e. contains ALP signatures. We then define the statistic
\begin{align}
\label{eq:ts-ml}
    \Pi(m_a,\gag|D)=1-p_{\rm ALP}(m_a,\gag|D)\,,
\end{align}
corresponding to the probability that dataset $D$ is not ALP-like according to the classifier at $(m_a,\gag)$. We compute a $\Pi$ distribution from 2000 simulated ALP-like datasets, and fit it to a Beta probability distribution function\footnote{\href{https://docs.scipy.org/doc/scipy-1.15.2/reference/generated/scipy.stats.beta.html}{https://docs.scipy.org/doc/scipy-1.15.2/reference/generated/scipy.stats.beta.html}}. We then compute the value of $\Pi$ corresponding to a given Asimov dataset without ALP features, denoted by $\Pi_A$, which we use to estimate the confidence level and Gaussian exclusion significance for the ALP hypothesis. 

As can be seen from the example distributions in Figure \ref{fig:example-ts-distribution-ml}, the classifier performance varies in different areas of the parameter space. In cases of good performance, the $\Pi$ distribution for ALP-like datasets is skewed towards 0, while the $\Pi_A$ value is close to 1, thus leading to exclusion (Figure \ref{fig:example-ts-distribution-ml-good}). On the other hand, where performance is worst, both the average $\Pi$ from the distribution and the $\Pi_A$ value from the Asimov dataset are found around 0.5, so that no exclusion can be set (Figure \ref{fig:example-ts-distribution-ml-bad}). The latter situation typically occurs for large $m_a$ and small $\gag$ values, where the ALP-like and ALP-less spectra are least distinguishable. 


\section{Results and conclusions}


As a reference case, we consider the 50-hour simulated observation of Mrk~501 in its baseline state. Preliminary results show that ALP-photon couplings down to $\gag\simeq8\times10^{-13}\,{\rm GeV}^{-1}$ in the mass range $m_a\in[10,100]\,{\rm neV}$ could be excluded at the $2\sigma$ confidence level by the likelihood-ratio approach. In addition, we find that the ML classifier method, introduced in the previous section, allows to extend the limits down to $\gag\simeq4\times10^{-13}\,{\rm GeV}^{-1}$ for $m_a\in[0.1,100]\,{\rm neV}$. In both cases, a significant portion of  previously unconstrained  parameter space is covered \cite{AxionLimits}. 


We point out, however, that these results were obtained from Asimov datasets. These, by definition, do not include Poissonian fluctuations on the photon counts in each energy bin. In a real analysis, this ``photon noise'' might be mistaken for a subtle ALP feature, especially at very low values of the $\gag$ coupling. As a consequence, our estimates could be weakened, as such values of $\gag$ could no longer be excluded. In addition, our results might be affected by various systematic effects, including magnetic field strength in the jet and assumed EBL and intrinsic spectral models. We plan to investigate all of these effects in a future paper, currently in preparation. 

We conclude with the remark that, among its many exciting prospects, CTAO will be an excellent tool to probe new fundamental physics. In particular, observations of blazars will have the capability to constrain a significant portion of the parameter space of ALPs beyond the regions previously excluded by experiment. In this contribution we have shown that, while the standard likelihood-ratio analysis technique already offers promising results in this regard, it is possible to apply an alternative method, based on the use of machine-learning classifiers, to further enhance the CTAO sensitivity to ALPs.

\acknowledgments
This work was conducted in the context of the CTAO DMEP working group. CTAO gratefully acknowledges financial support from the agencies and organizations listed at \href{https://www.ctao.org/for-scientists/library/acknowledgments/}{https://www.ctao.org/for-scientists/library/acknowledgments/}. We would like to thank the computing centers that provided resources for the IRF generation, listed at \href{https://zenodo.org/records/5499840}{https://zenodo.org/records/5499840}.

\bibliographystyle{JHEP}
\bibliography{biblio.bib}

\end{document}